\documentclass[graybox]{svmult}

\usepackage{mathptmx}       
\usepackage{helvet}         
\usepackage{courier}        
\usepackage{type1cm}        
\usepackage{makeidx}         
\usepackage{multicol}        
\usepackage[bottom]{footmisc}

\usepackage{cite}
\usepackage[pdftex]{graphicx}
\DeclareGraphicsExtensions{.eps,.pdf,.bmp,.png,.mps,.jpg}
\usepackage{subfigure}
\usepackage{amssymb}

\makeindex             

\begin{document}

\title*{Congestion Dynamics in Pedestrian Single-File Motion}
\author{Verena Ziemer, Armin Seyfried and Andreas Schadschneider}
\institute{Verena Ziemer \at Computer Simulation for Fire Safety and Pedestrian Traffic, Bergische Universit\"at Wuppertal, 42285 Wuppertal, Germany
\\ \email{ziemer@uni-wuppertal.de}
\and Armin Seyfried \at J\"ulich Supercomputing Centre, Forschungszentrum J\"ulich GmbH, 52425 J\"ulich, Germany
\\ \email{a.seyfried@fz-juelich.de}
\and Andreas Schadschneider \at Institut f\"ur Theoretische Physik, Universit\"at zu K\"oln, 50937 K\"oln, Germany
\\ \email{as@thp.uni-koeln.de}}

\maketitle

\abstract{This article considers execution and analysis of laboratory experiments of pedestrians moving in a quasi-one-dimensional system with periodic boundary conditions.
To analyze characteristics of jams in the system we aim to use the whole experimental setup as the measurement area.
Thus the trajectories are transformed to a new coordinate system.
We show that the trajectory data from the straight and curved parts are comparable and assume that the distributions of the residuals come from the same continuous distribution.
Regarding the trajectories of the entire setup, the creation of stop-and-go waves in pedestrian traffic can be investigated and\linebreak described.}

\section{Introduction}
\label{sec:introduction}
In recent years many research groups have executed experiments with pedestrians, see e.g. \cite{Hoogendoorn2003,Seyfried2005,Chattaraj2009,Jelic2012,Yanagisawa2012}.
One phenomenon which can be observed is a stop-and-go wave well known from vehicular traffic, e.g., \cite{Chowdhury2000}.
It is visible in the simplest system with a one-dimensional movement of pedestrians along a line with closed boundary conditions, e.g., for a ring \cite{Seyfried2005} and for a circle \cite{Jelic2012}.

In the following we summarize previous work on one-dimensional pedestrian movements where stop-and-go waves were analyzed.
\cite{Seyfried2005} executed the first one-dimensional experiments in 2005.
For the first time high densities were examined in \cite{Portz2010} and the existence of stop-and-go waves in pedestrian traffic was shown.
In \cite{Portz2010} also an adaptive velocity model with reaction time was proposed, which is able to reproduce qualitatively similar stop-and-go waves as they are observable in the experimental data, see \cite{Seyfried2010}.
A coexistence of two differing speed phases in this data was shown in \cite{Portz2011} by analyzing the fundamental diagram in specific density regions.
A model that reproduces this coexistence was also proposed.
The same experimental data was taken in \cite{Eilhardt2014} to test a stochastic headway dependent velocity model.
\cite{Jelic2012} compared data from a circle with data from a straight line.
Beside the free flow and congested regime of the fundamental diagram they found a third regime named weakly constrained regime.
This regime lies between the free and the strongly congested regime.

In \cite{Seyfried2010}, the measurement section of the nearly $27$ m corridor with closed boundary conditions covers $4$ m only.
This restriction makes it impossible to test models concerning number and dimension of a stop-and-go wave because of lack of data.
A second restriction is the duration of an experiment run.
In experiments, the duration may be too short to reach a stable state.
That is why the lifetime of a stop-and-go wave, its length in space and the number of stop-and-go waves in the system cannot be observed.

In this article we want to handle this problem.
Experiments where the whole system was observed are analyzed.
Furthermore, we introduce a methodology to analyze characteristics of stop-and-go waves.

\section{Experiments}
\label{sec:experiments}
Laboratory experiments were performed within the framework of the project BaSiGo 
in June 2013.
The aim of those experiments is a better understanding of pedestrian dynamics in critical crowded states, e.g., in front of an entrance to a concert hall, thus in high densities.
It included large-scale experiments in various scenarios with up to 1000 pedestrians per run during five days.
The probands were mostly students ($55\%$ male and $45\%$ female) at the age of 18 to 72 years, $25$ years on average, with an average height of $1.71$~m and recieved 50 Euro per day.
$13\%$ of them were living in suburbs or rural areas and $68\%$ in cities or major cities.

\subsection{Experimental setup}
\label{ssec:setup}
One-dimensional laboratory experiments for pedestrians in a system with periodic boundary conditions were executed.
The setup is a ring corridor with a circumference about $C=26.5$~m ($r_0=3$~m, $l=4$~m).
Six runs were performed with\linebreak different number of participants ($N=15,\,30,\,47,\,55,\,52,\,59$) so that the global density ($\rho_g=N/C$) ranges from $0.57$ to $2.27$~Ped/m.
For each run, the pedestrians initially are arranged uniformly in the setup.
Then there are two commands, the first to start walking with normal speed in the corridor and the second for stopping.

\subsection{Data collection}
\label{ssec:collection}
\begin{figure}[t]
\centering
\subfigure{
\includegraphics[width=0.45\linewidth,trim=0pt 0pt 0pt 15pt,clip]{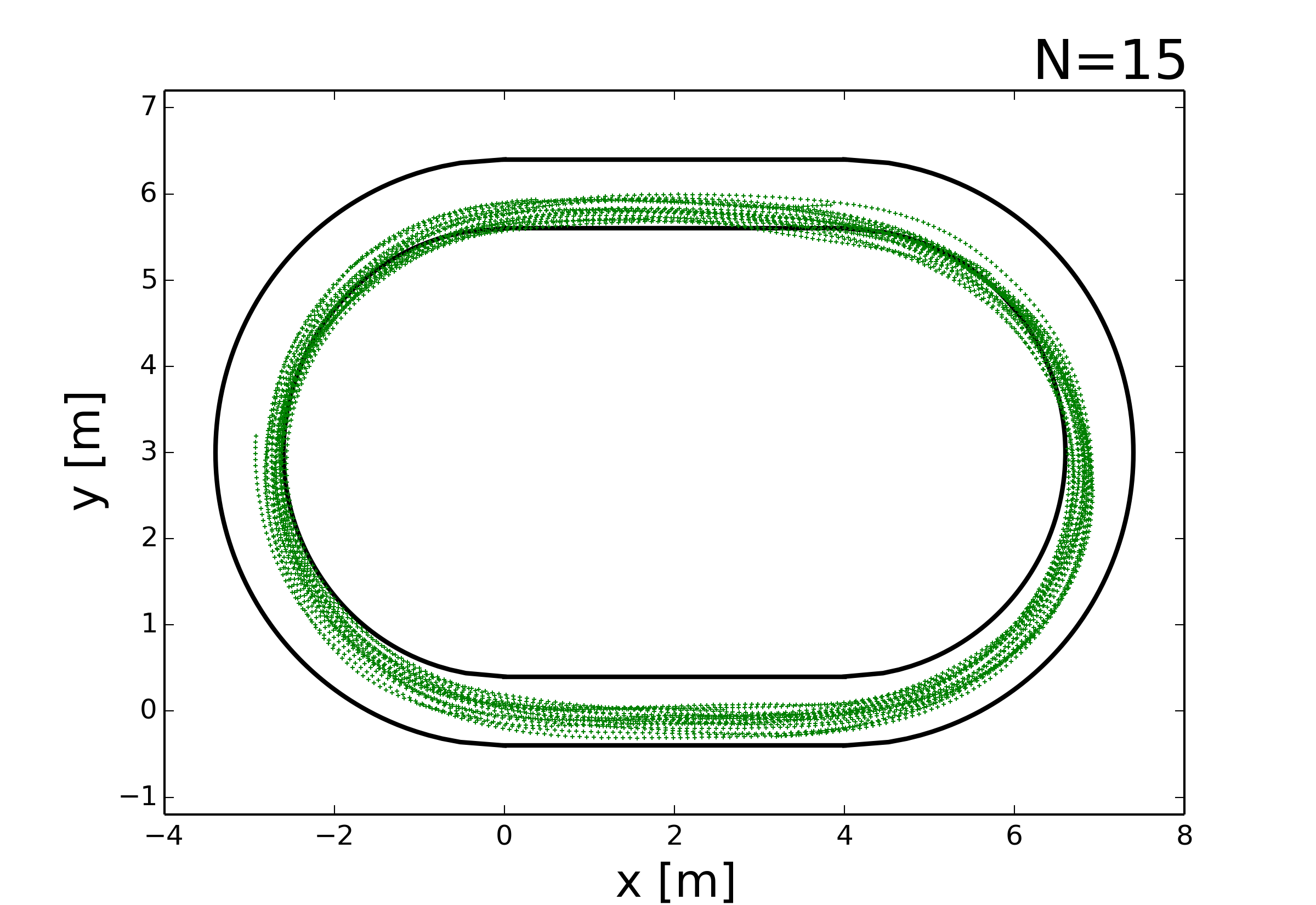}
\label{subfig:ring15}
}
\subfigure{
\includegraphics[width=0.45\linewidth,trim=0pt 0pt 0pt 15pt,clip]{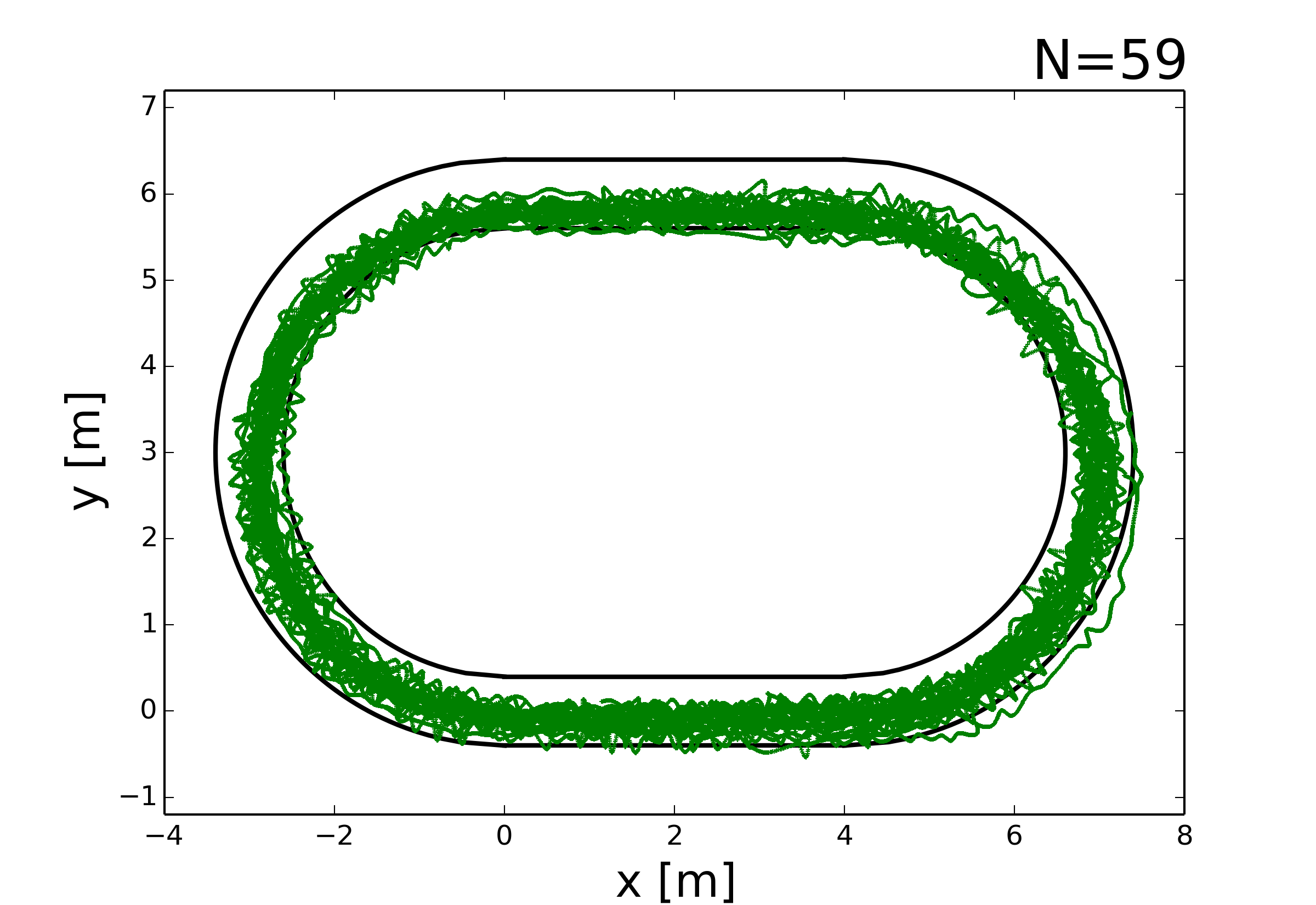}
\label{subfig:ring59}
}
\caption[Trajectories in oval]{Extracted trajectories for $N=15$ and $N=59$}
\label{fig:ring}
\end{figure}

To enable the automatic tracking of the trajectories, the pedestrians wear hats with a marker.
For each run the whole setup was recorded from the top video camera with a frame rate of $16$~fps.
From those video recordings pedestrian trajectories who are describing the movement of the  heads of the pedestrians were extracted.
For more details to this method we refer to\cite{Mehner2015} in these proceedings.
Fig.~\ref{fig:ring} shows the trajectories for the runs with $N=15$ and $N=59$ which leads to the global density $0.57$~Ped/m and $2.27$~Ped/m, respectively.

\subsection{Data preparation}
\label{ssec:2.3}
\begin{figure}[t]
\centering
\subfigure{
\includegraphics[width=0.45\linewidth,trim=0pt 0pt 0pt 20pt,clip]{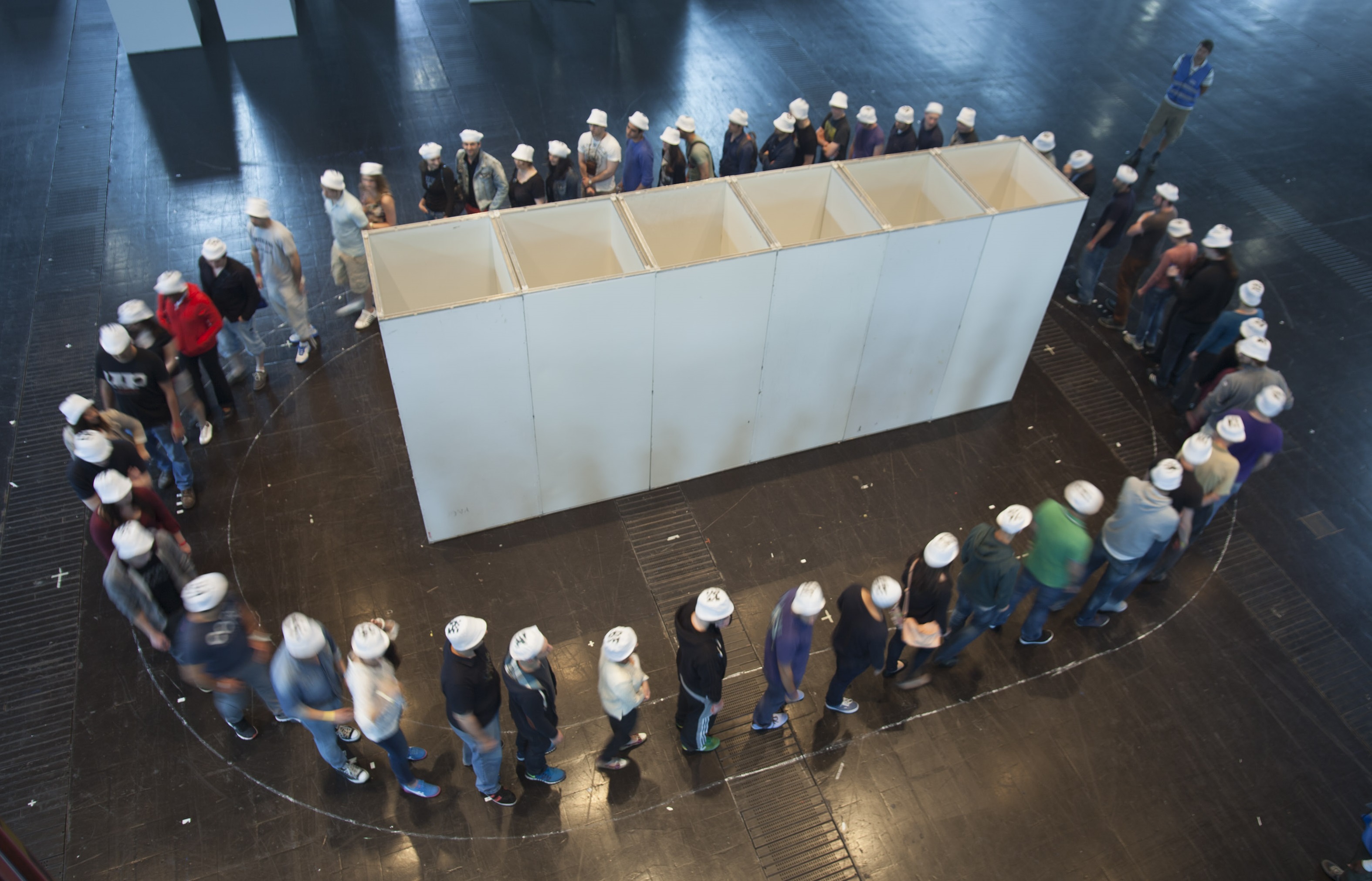}
}
\subfigure{
\includegraphics[width=0.45\linewidth,trim=0pt 0pt 0pt 30pt,clip]{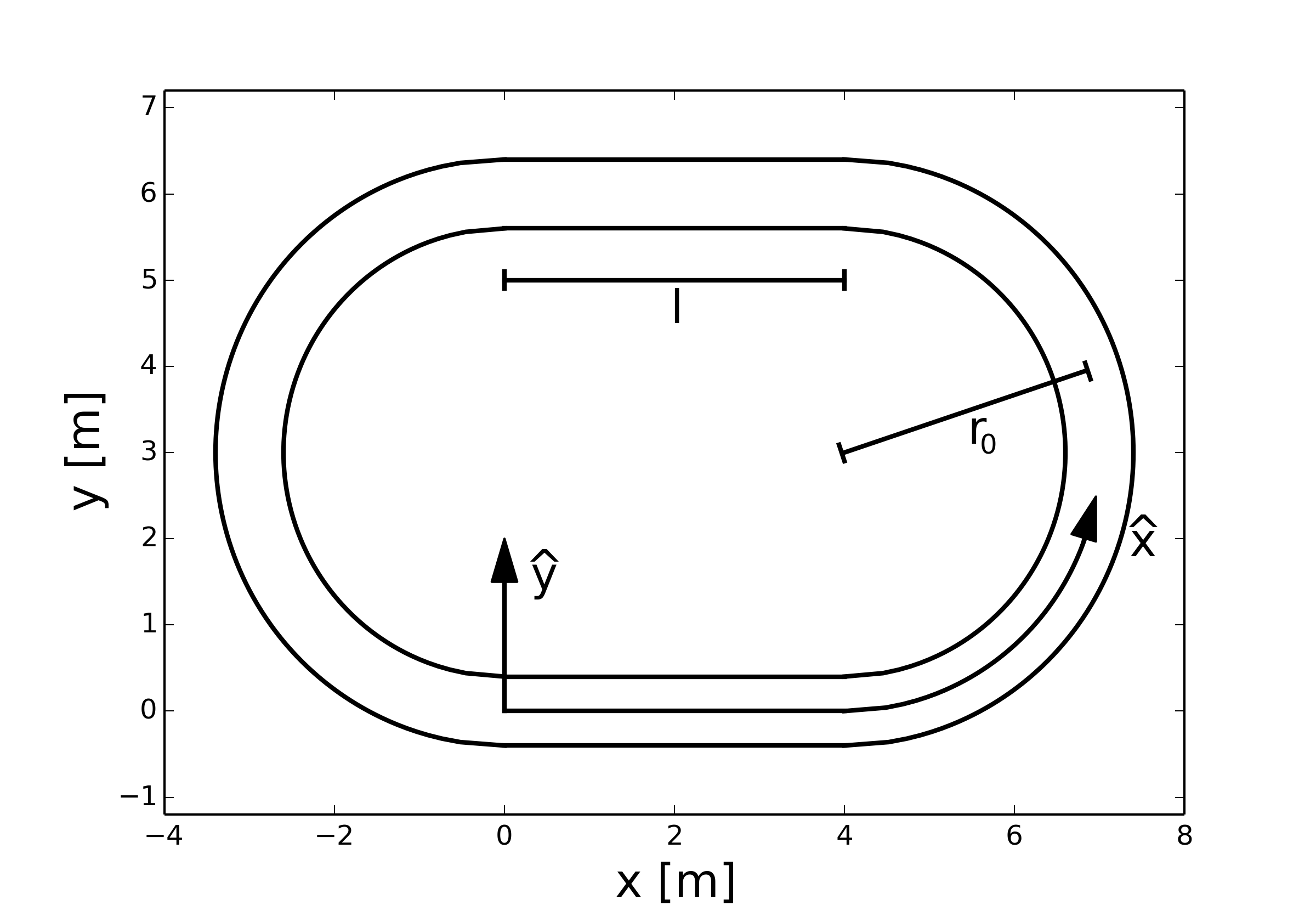}
}
\caption[Transformation]{Left: Photo from the experiment by Marc Strunz. Right: New coordinate system $\hat{x}$, $\hat{y}$}
\label{fig:transformation}
\end{figure}

We want to study one-dimensional characteristics of stop-and-go waves.
Therefore, we introduce a coordinate $\hat{x}$ along the ring which corresponds to the distance from the origin measured along the middle line of the corridor.
It corresponds to the walking distance of a person staying always in the middle.
The coordinate $\hat{y}$ is always perpendicular to the $\hat{x}$-axis.
It measures the deviation from the center of the corridor, see Fig.~\ref{fig:transformation}.
We call $\hat{x}$ the main position and $\hat{y}$ the orientated distance.

The transformation $\lambda: \mathbb{R}^2\rightarrow\mathbb{R}^2$, 
$\left(\begin{array}{c} x \\ y \end{array}\right)\mapsto
\left(\begin{array}{c} \hat{x} \\ \hat{y} \end{array}\right)$
is described by:

\begin{eqnarray}
\hat{y}=\left\lbrace \begin{array}{ll}
\sqrt{x^2+(y-r_0)^2} -r_0		&\quad x<0,\\
\sqrt{(y-r_0)^2} -r_0	&\quad 0\leq x\leq l,\\
\sqrt{(x-l)^2+(y-r_0)^2} -r_0		&\quad x>l,
\end{array}
\right.
\label{eq:02}
\\
\hat{x}=\left\lbrace \begin{array}{ll}
2l+r_0\pi+r_0\arccos(\frac{r_0-y}{\sqrt{(x-l)^2+(y-r_0)^2}}) &\quad x<0,\\
x 				&\quad 0\leq x\leq l,~ y<r_0,\\
2l+r_0\pi-x 	&\quad 0\leq x\leq l,~ y\geq r_0,\\
l+r_0\arccos(\frac{r_0-y}{\sqrt{x^2+(y-r_0)^2}}) 	&\quad x>l.
\end{array}
\right.
\label{eq:01}
\end{eqnarray}

\begin{figure}[t]
\centering
\subfigure{
\includegraphics[width=0.45\linewidth,trim=0pt 0pt 0pt 15pt,clip]{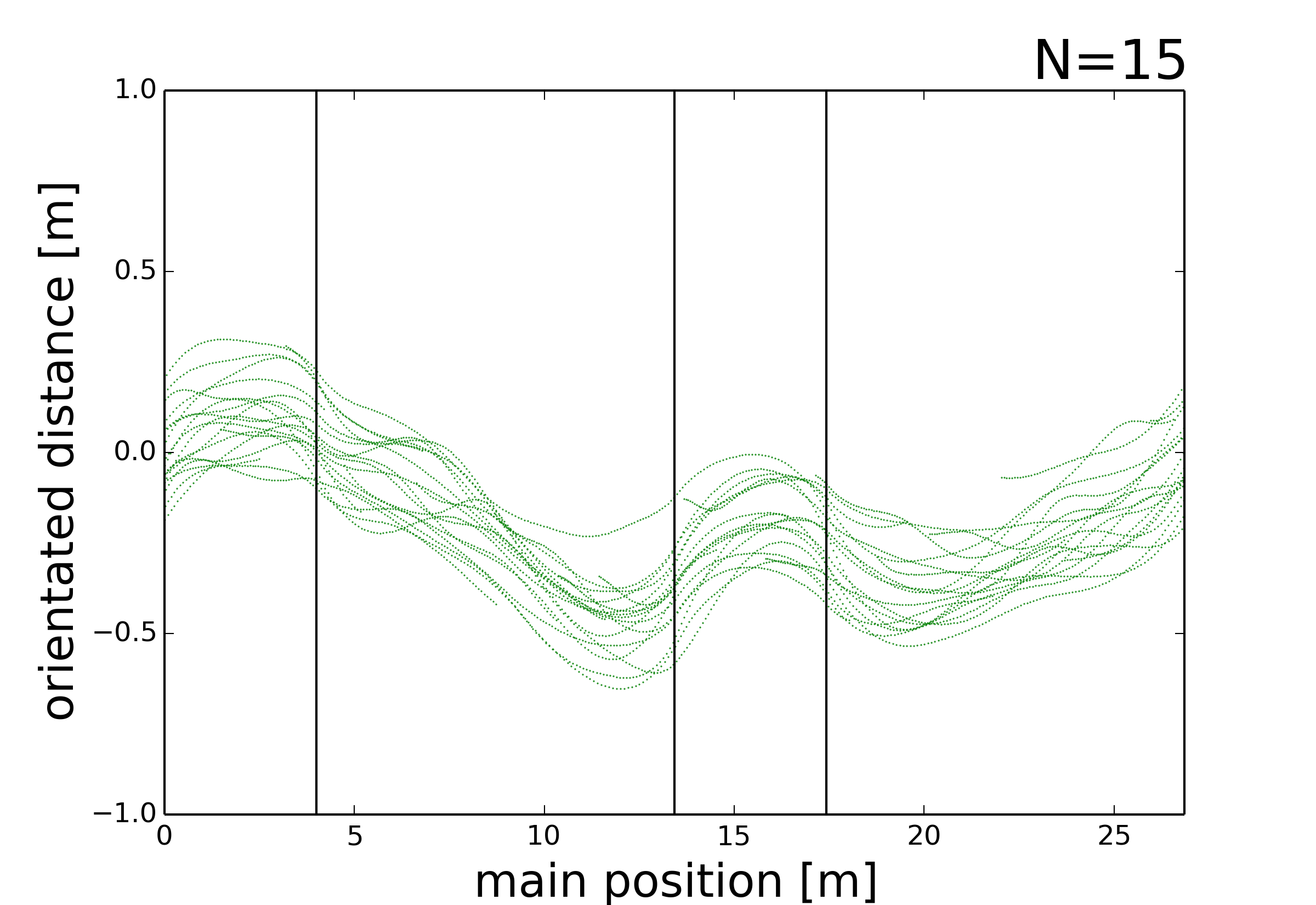}
\label{subfig:trajectory15}
}
\subfigure{
\includegraphics[width=0.45\linewidth,trim=0pt 0pt 0pt 15pt,clip]{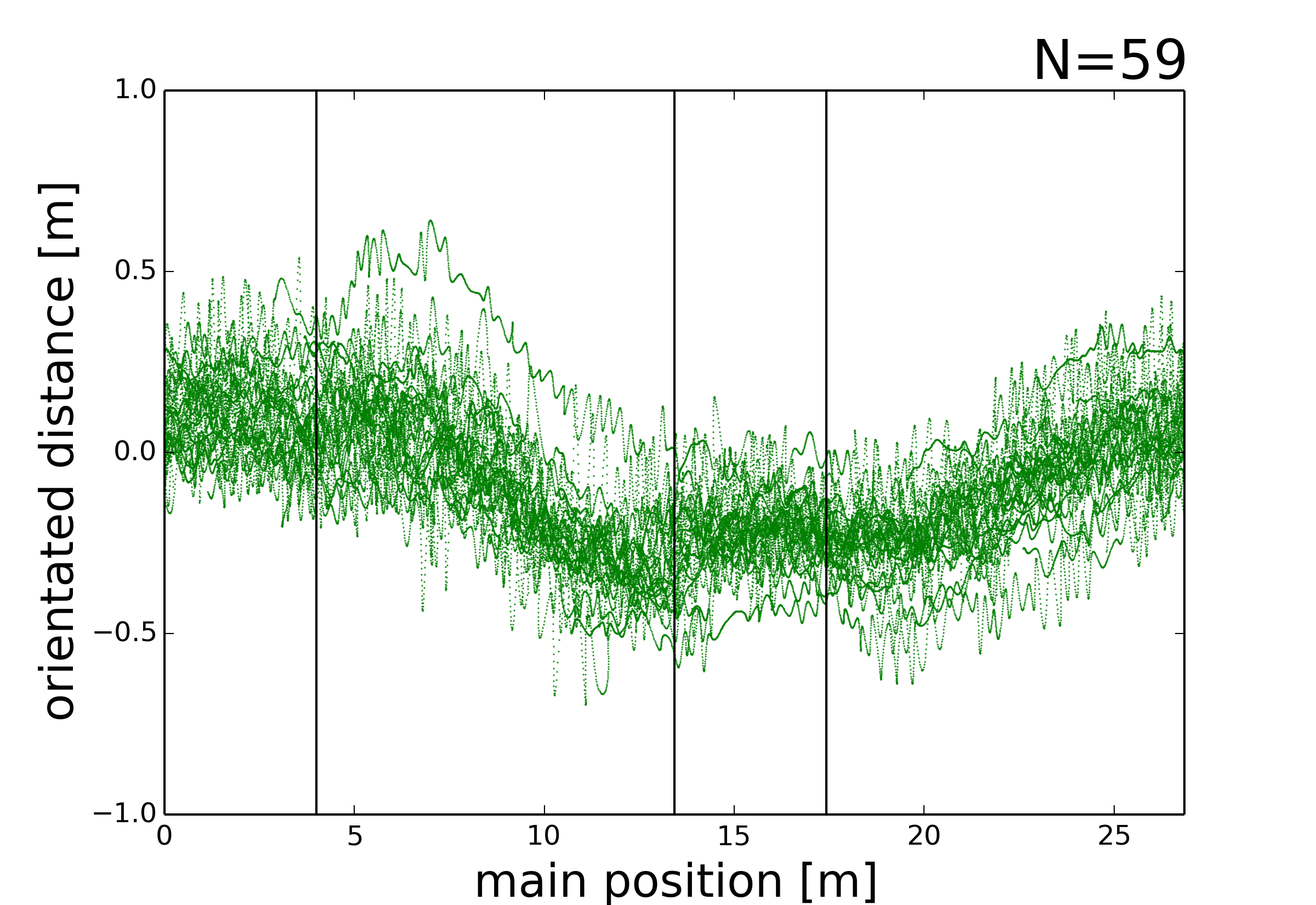}
\label{subfig:trajectory59}
}
\caption[Run with 15 persons]{Adjusted trajectories for $N=15$ and $N=59$ in the new coordinate system}
\label{fig:trajectory}
\end{figure}

Fig.~\ref{fig:trajectory} shows the trajectories of adjusted data in the new coordinate system for two runs.
The straight lines in the ring correlate to the intervals $[0,4]$~m and $[14.42,18.42]$~m of the main position.
Just as well the two other intervals correlate to the two curved parts.
In the following, we will only examine the main position, described by Eq.~\ref{eq:01} and omit the orientated distance.
That means that all further calculations come from this one-dimensional main position.

\section{Results}
\label{sec:results}
In this section we test whether the fundamental diagram has different properties in the straight and curved part of the corridor, i.e. whether it is influenced by the curvature.

\subsection{Curvature-dependence of the fundamental diagram}
\label{curvature}
\begin{figure}[t]
\centering
\subfigure{
\includegraphics[width=0.45\linewidth,trim=0pt 0pt 0pt 30pt,clip]{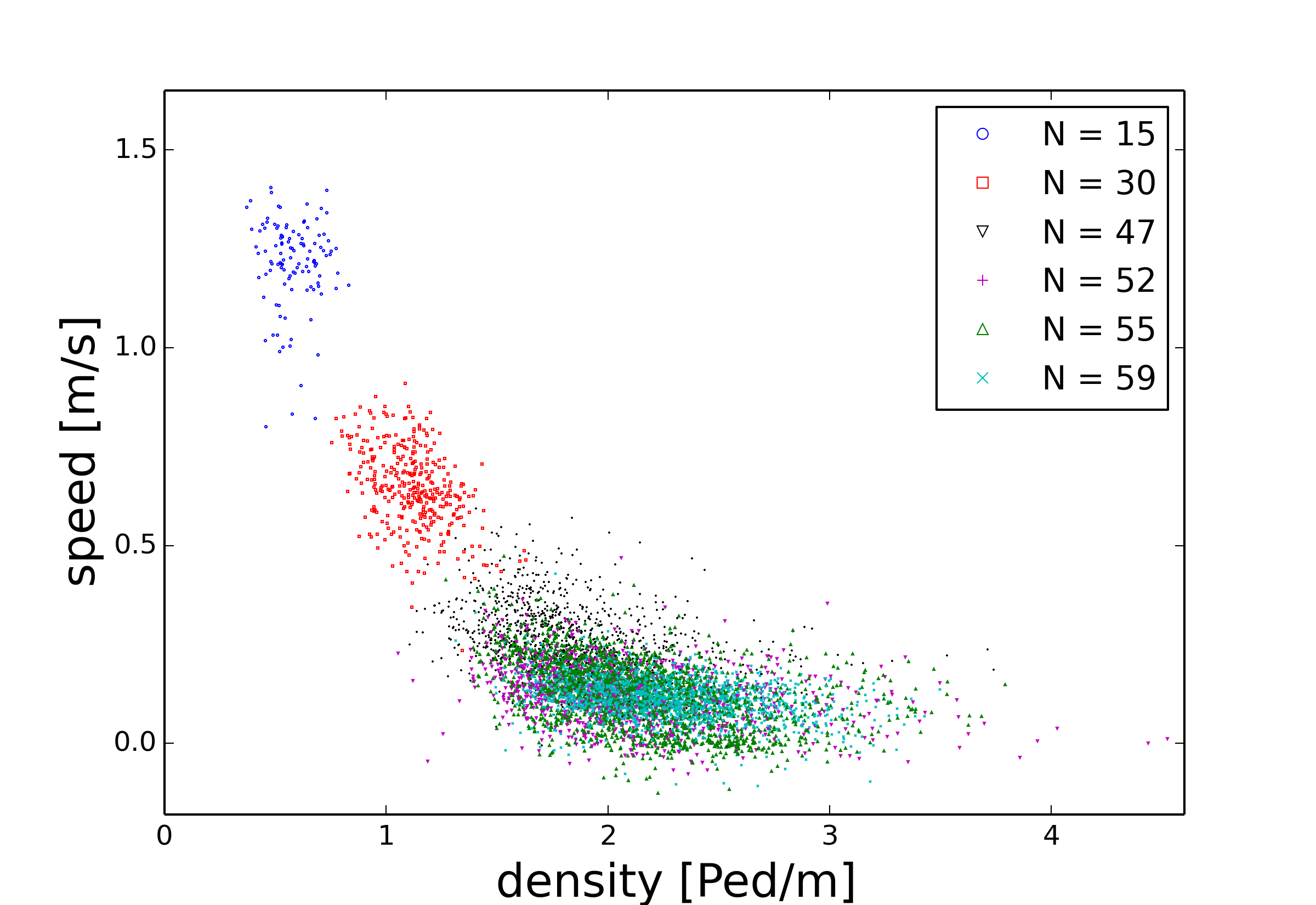}
\label{subfig:fd_line}
}
\subfigure{
\includegraphics[width=0.45\linewidth,trim=0pt 0pt 0pt 30pt,clip]{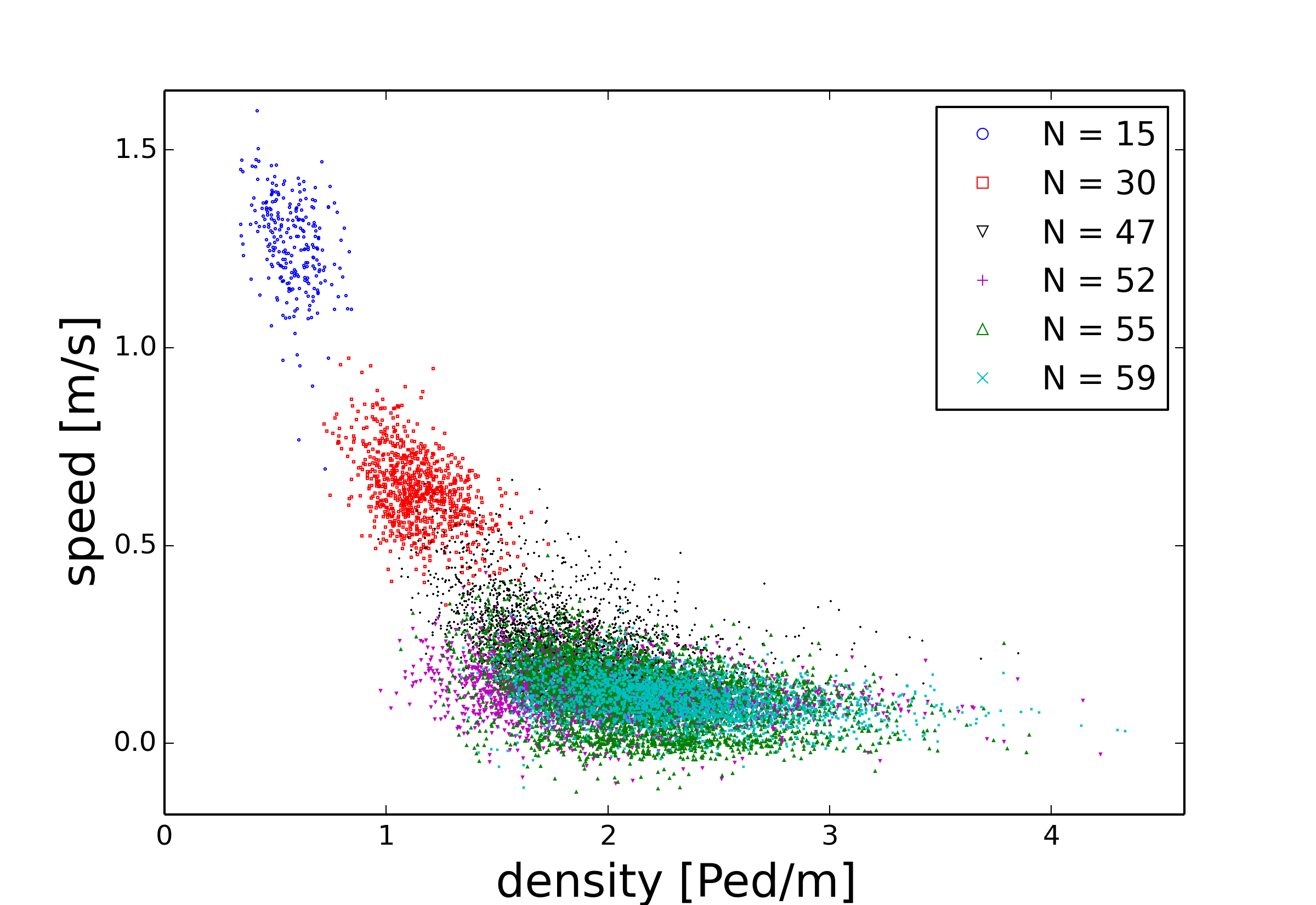}
\label{subfig:fd_circle}
}
\caption[FD]{The fundamental diagram of single file pedestrian motion. Left: Data from the straight part. Right: Data from the curved part}
\label{fig:fd_line_circle}
\end{figure}

The ring corridor, our experimental setup, has four parts, twice the straight line and twice the half circle.
Fig.~\ref{fig:fd_line_circle} shows the Voronoi-based fundamental diagram for pedestrians moving in the straight line and in the curve.
To distinguish between the six runs, each run has a separate symbol and color.

The speed of pedestrian $i$ at time $t$ is calculated by the position difference over half a second divided by the time $v_{i}(t)=(\hat{x}_{i}(t+\Delta t)-\hat{x}_{i}(t-\Delta t))/(2\Delta t)$ with\linebreak $\Delta t = 0.25$~s.
The density is then $\rho_{i}(t)={d_{i}}^{-1}$, where the Voronoi space $d_{i}$ of pedestrian $i$ is half the distance between the two neighbors $i-1$ and $i+1$,\linebreak $d_{i}=(\hat{x}_{i-1}-\hat{x}_{i+1})/2$.

For analyzing the characteristics of stop-and-go waves in the whole system we want to combine the data of both parts.
First of all we test the comparability of the fundamental diagram for the straight line and the curve.
That means we study the distribution for both parts.
The Kolmogorov-Smirnov test gives an indication whether two data clouds have the same distribution or not.
A precondition for this test is that the data has to be independent, which can be shown with the autocorrelation.

\begin{figure}[t]
\includegraphics[width=0.95\linewidth,trim=0pt 0pt 0pt 0pt,clip]{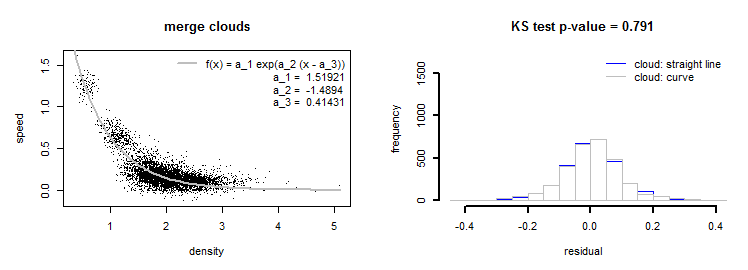}
\label{fig:pvalue}
\caption[Kolmogorov-Smirnov]{Left: Fundamental diagram of single file pedestrian motion for 6000 data points and its fitted exponential function. Right: Result of the Kolmogorov-Smirnov test}
\label{fig:Kolgomorov}
\end{figure}

We use two clouds with 3000 independent data points of each category, see Fig.~\ref{fig:Kolgomorov} on the left side.
The autocorrelation was tested for those data points to ensure that the observations are independent.
An exponential function $f$ for the shape of the speed according to the density is fitted by least square.
The distributions of the residuals for the clouds are tested with the Kolmogorov-Smirnov test.
The null hypothesis is H0: 'the distribution is the same for the two samples'.
The frequency of residuals for both parts is shown in Fig.~\ref{fig:Kolgomorov}, right.
With $p=0.791$ we can clearly not reject the assumption that the distributions of the residuals come from the same continuous distribution.

With the Kolmogorov-Smirnov test we have shown that the density speed relation in the straight line is subject to the distribution of the density speed relation in the curve.
That means that curvature effects on the fundamental diagram can be neglected.

\subsection{Visualization of stop-and-go waves}
\label{ssec:3.2}
\begin{figure}[!t]
\centering
\subfigure{
\includegraphics[width=0.45\linewidth,trim=0pt 0pt 0pt 30pt,clip]{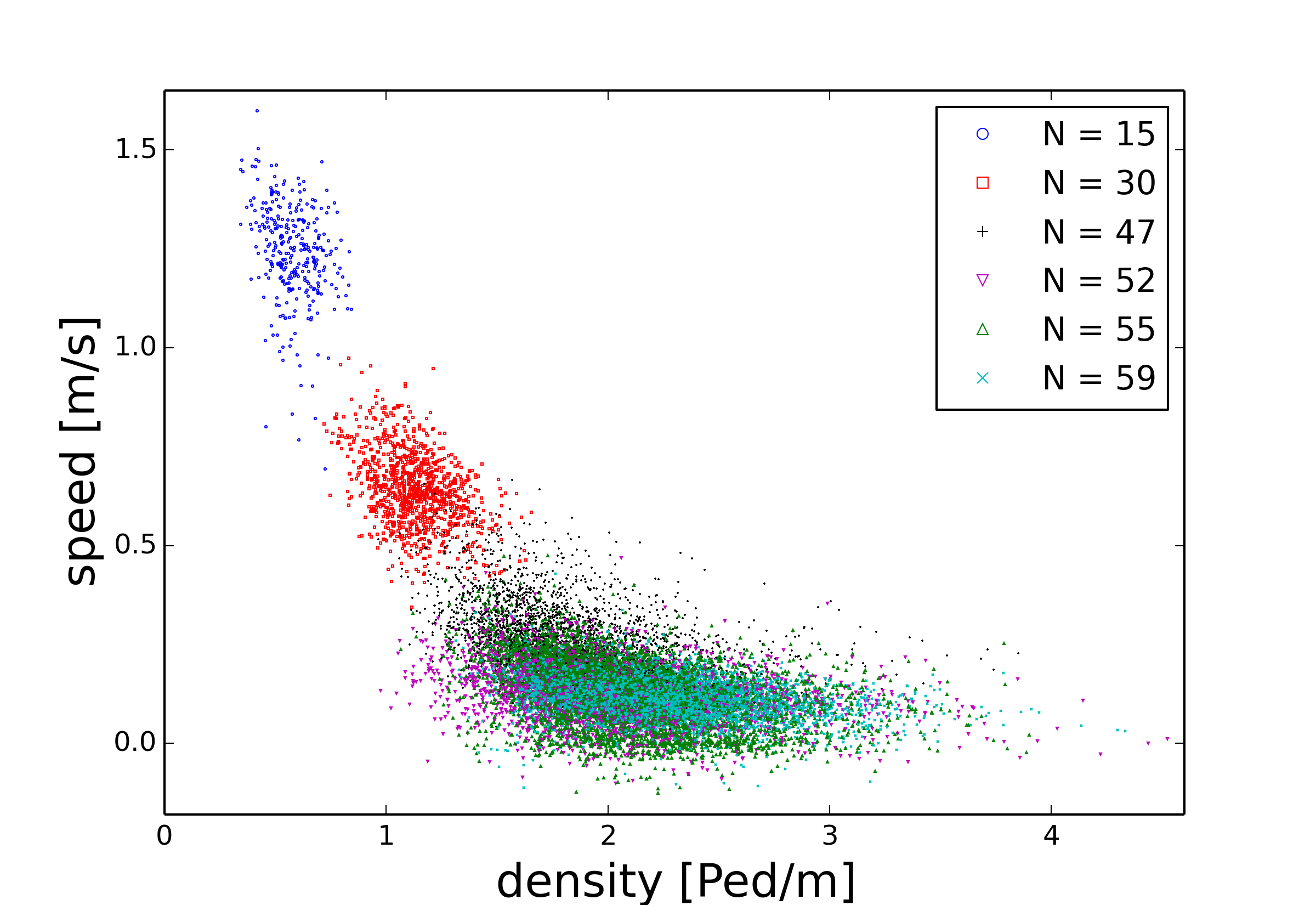}
\label{subfig:bin1}
}
\subfigure{
\includegraphics[width=0.45\linewidth,trim=0pt 0pt 0pt 30pt,clip]{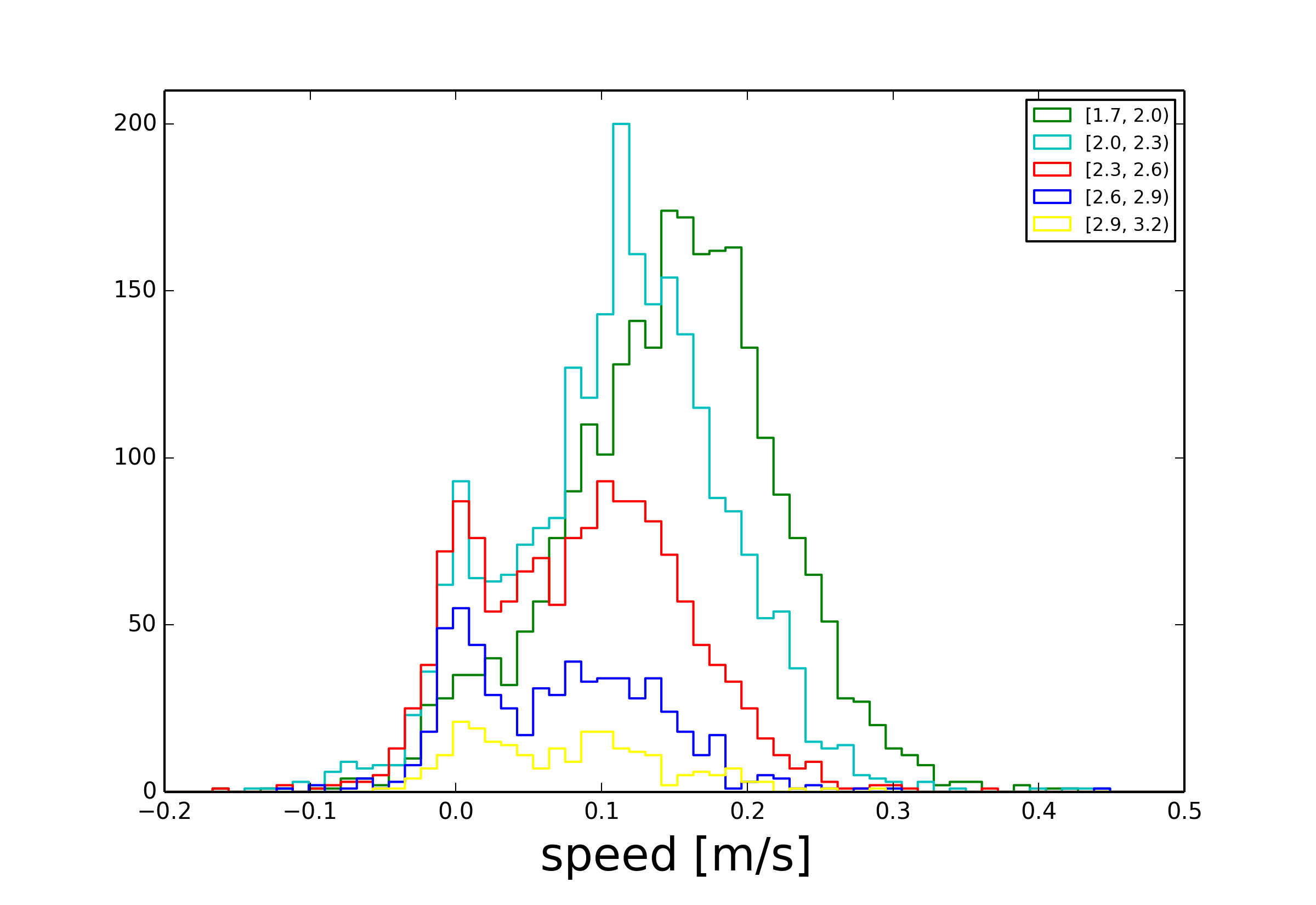}
\label{subfig:bin2}
}
\caption[Frequency]{Left: Fundamental diagram of single file pedestrian motion. Right: Frequency of speed for certain density regions, $N=55$}
\label{fig:coexistence}
\end{figure}

To visualize stop-and-go waves, we use the knowledge that there is a coexistence of two separate speed phases.
The fundamental diagram for the whole system for all six runs is shown in Fig.~\ref{fig:coexistence} on the left side.
By analyzing the frequency of speeds in specific density regions the coexistence of two differing speed regimes can be identified.
One peak is around the speed $0.12$~m/s and the other one around $0$~m/s.
Negative speeds result from the swaying of heads of standing pedestrians.

In order to distinguish both phases, we introduce a stop speed $v_{stop}$.
Standing pedestrians have a speed lower than this stop speed and moving pedestrians have a speed which is higher than the stop speed.
We set the stop speed to $v_{stop}=0.05$~m/s, which appears to be a reasonable value for our experiments.
We have to note, that this value is not fixed and there might be better ones for other experiments.

\subsection{Stop-and-go waves}
\label{ssec:3.3}
\begin{figure}[!t]
\centering
\subfigure{
\includegraphics[width=0.45\linewidth,trim=0pt 0pt 0pt 15pt,clip]{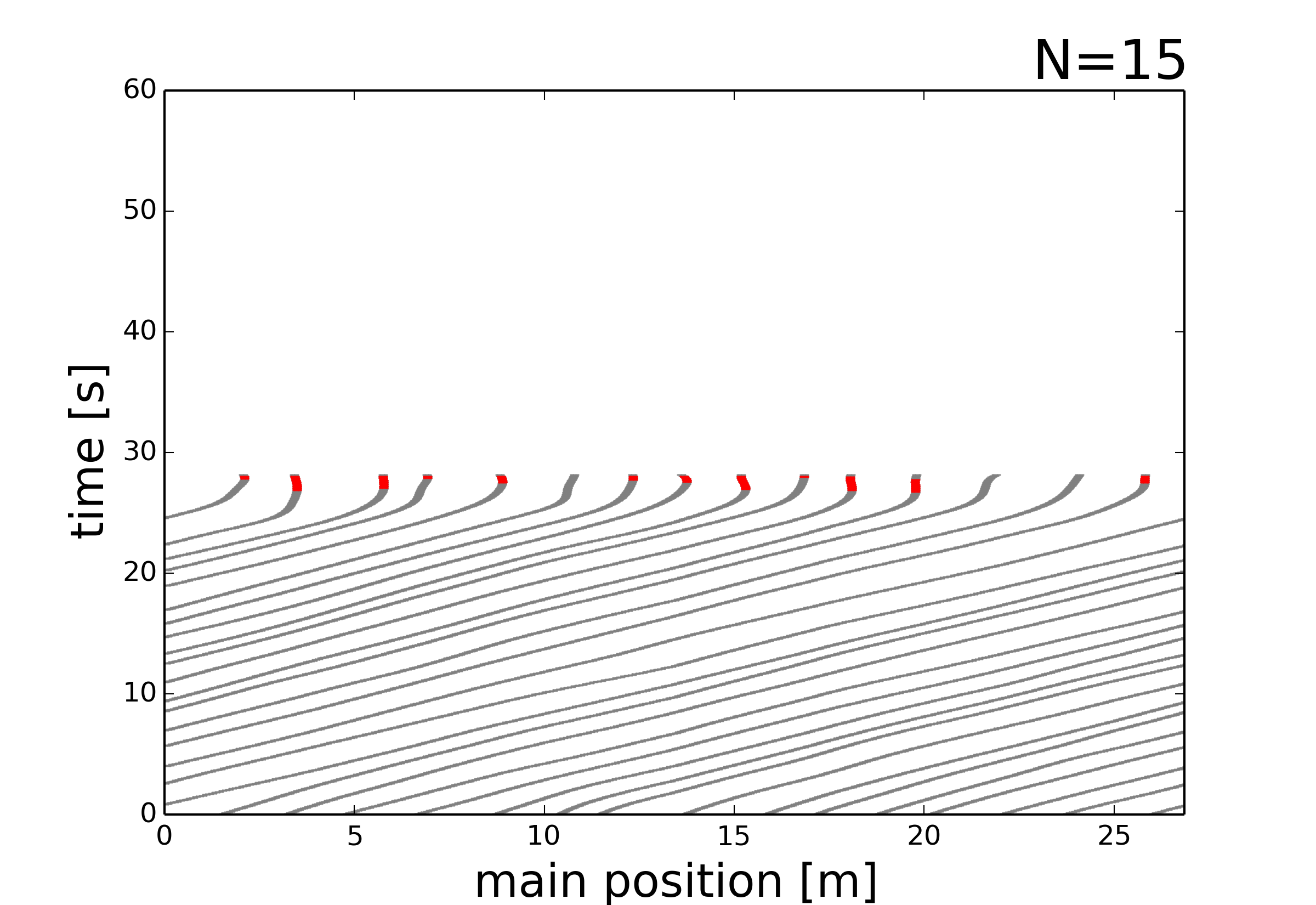}
\label{subfig:time15}
}
\subfigure{
\includegraphics[width=0.45\linewidth,trim=0pt 0pt 0pt 15pt,clip]{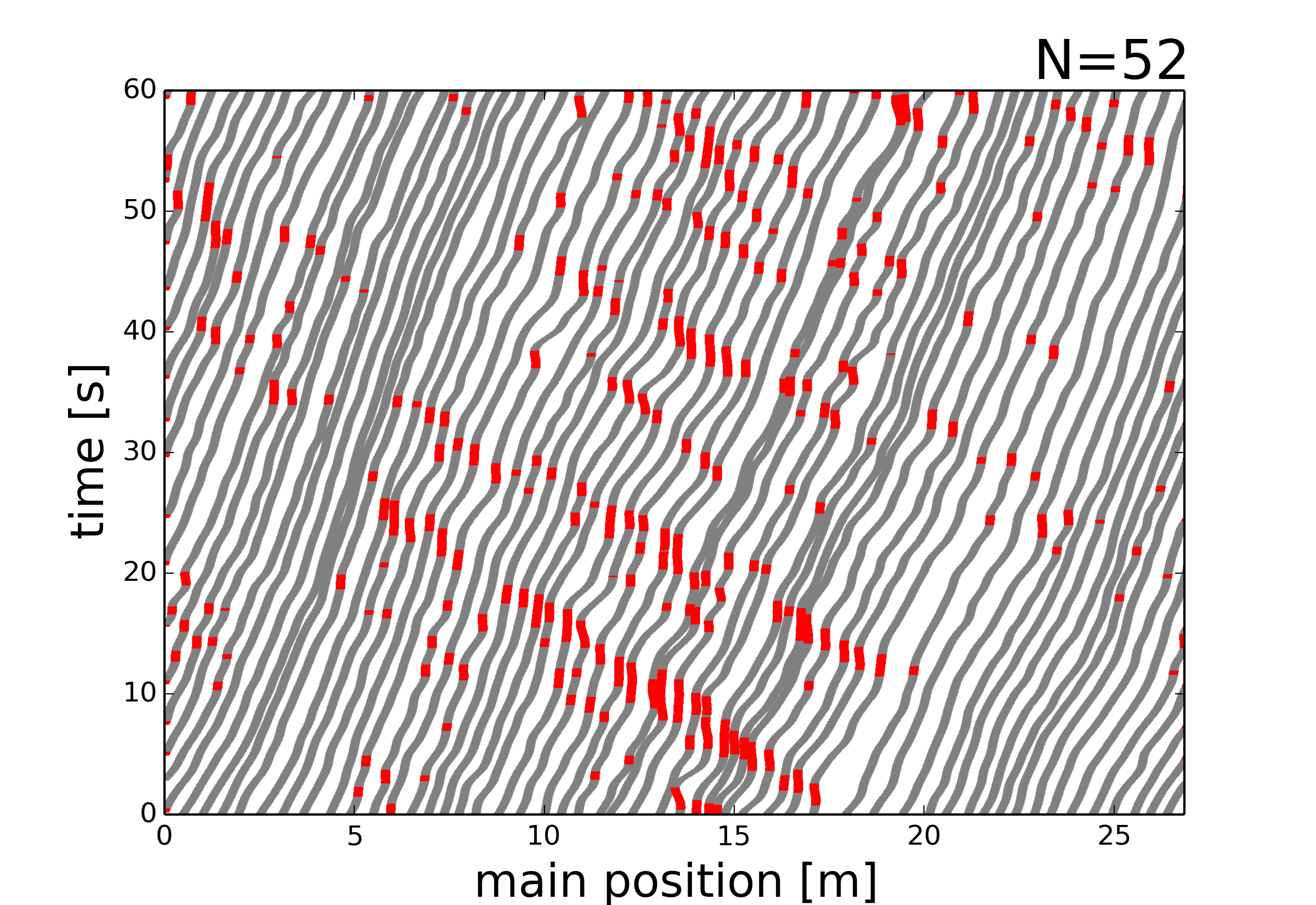}
\label{subfig:time52}
}
\subfigure{
\includegraphics[width=0.45\linewidth,trim=0pt 0pt 0pt 15pt,clip]{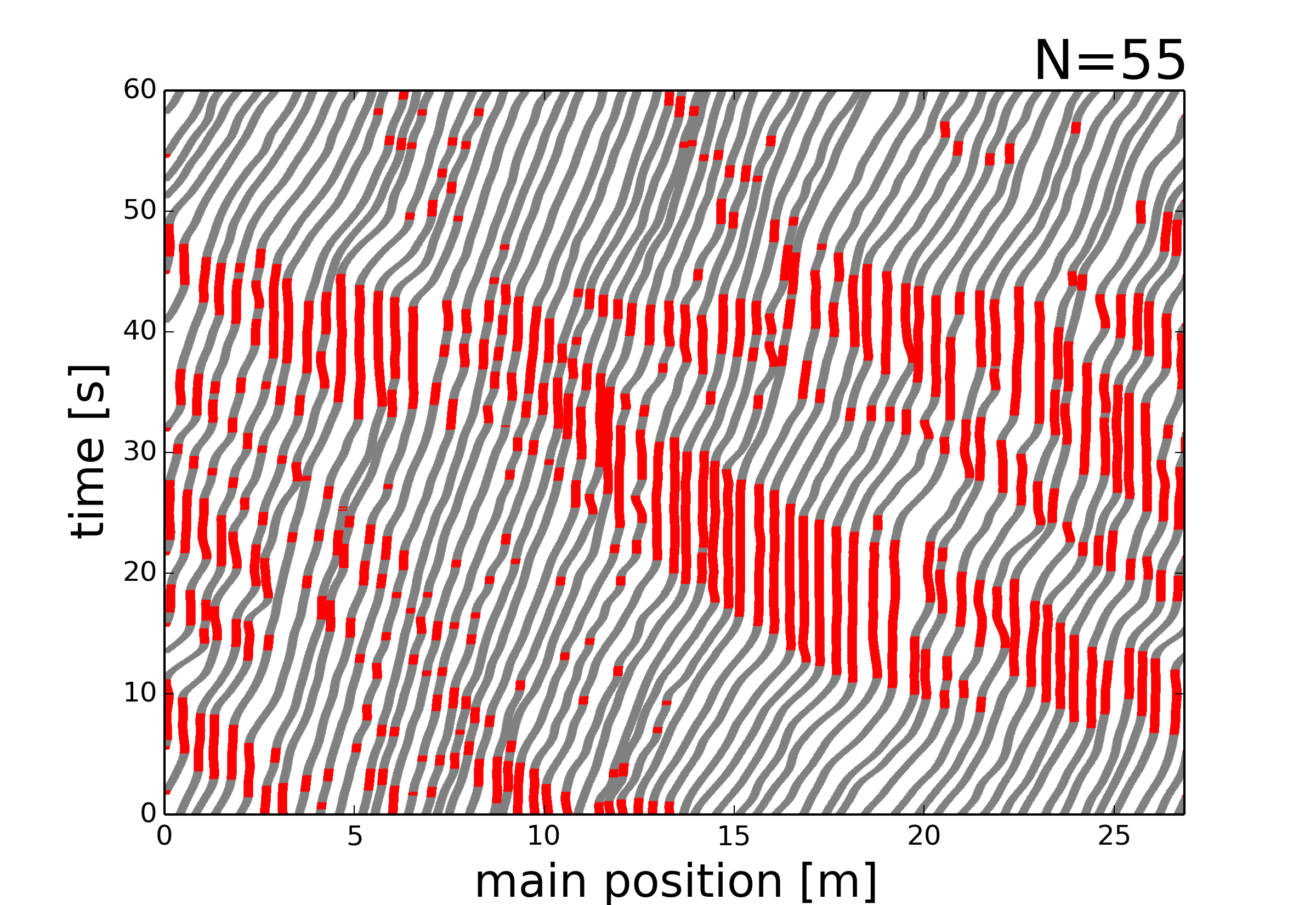}
\label{subfig:time55}
}
\subfigure{
\includegraphics[width=0.45\linewidth,trim=0pt 0pt 0pt 15pt,clip]{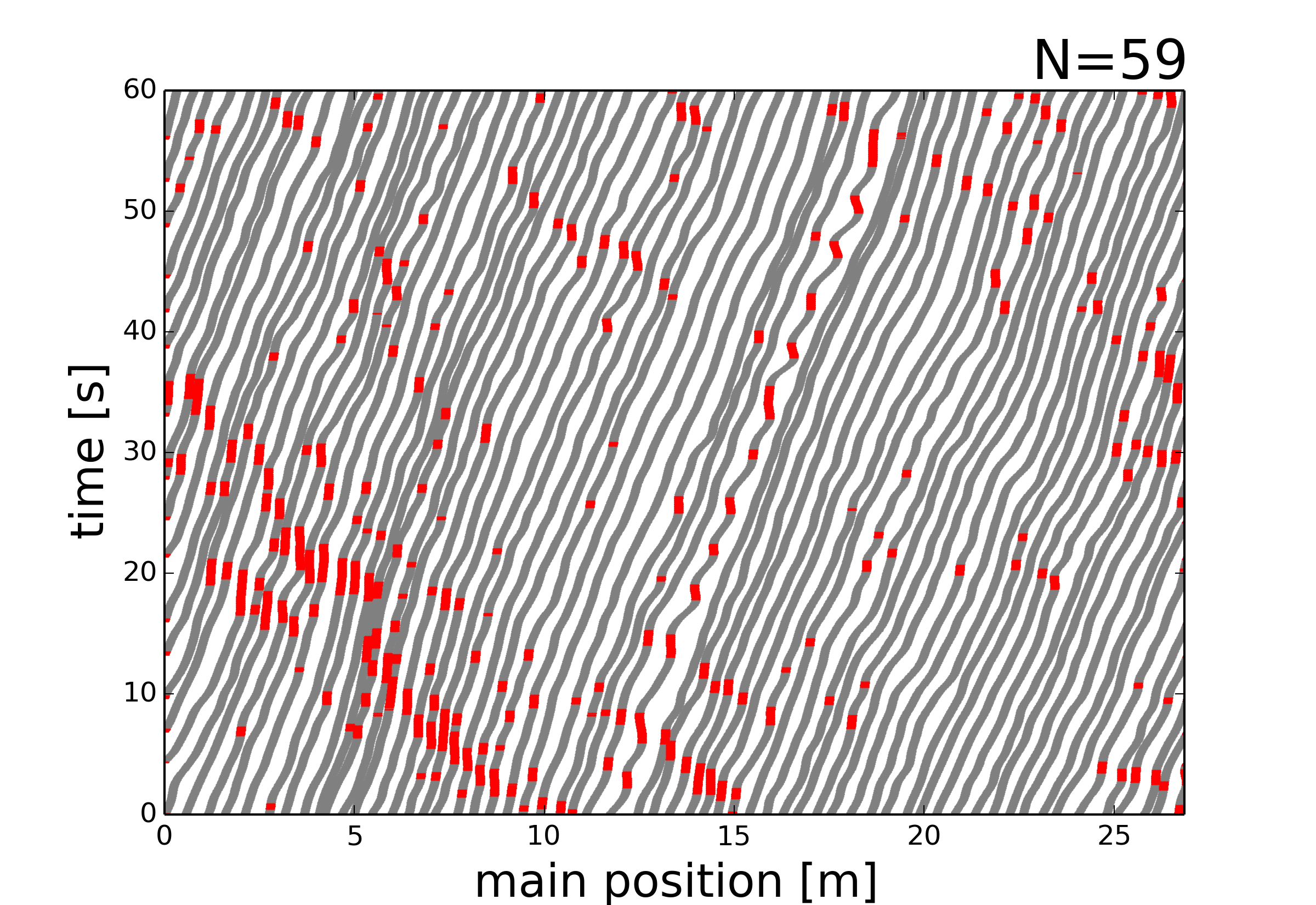}
\label{subfig:time59}
}
\caption[ORT-Zeit-Kurve]{Trajectories of the main positions depending on time, $N=15,\,52,\,55,\,59$.
A red dot indicates a pedestrian in a congestion, that means he has a speed not higher than the stop speed}
\label{fig:time1}
\end{figure}

We define a stop-and-go wave by a consecutive sequence of one or more standing pedestrians.
Fig.~\ref{fig:time1} shows the main positions of the pedestrians for one minute for four different runs.
The plotted positions have a length of $0.2$~m to represent a body length.
Pedestrians in a stop wave are marked in red.
With 15 pedestrians in the system, no stopping occurs whereas 52, 55 or 59 pedestrians generate stop-and-go waves.
Eye-catching is the run with 55 pedestrians.
Long and big stop-and-go waves can be observed.
The duration of the experiment run was too short to judge whether the regarded time interval represents a stable state.

A stop wave can be characterized by the number of pedestrians in this region at a certain time, by the length in space at a certain time and by the duration at a certain time and position.
The length of a stop wave can be described by the number of standing pedestrians in a line or the distance between the first standing pedestrian and the last one.
Higher densities lead to stop waves with more standing pedestrians than lower densities whereas the average length of a stop wave gets shorter.
The duration of a stop wave can be described by the time while the first pedestrian in this stop wave is standing and the last standing pedestrian starts to move. 

In the run $N=52$, all stop waves have nearly the same speed around $-0.5$~m/s.
The run $N=59$ has lower speeds of stop waves around $-0.32$~m/s.

\section{Conclusion}
\label{sec:4}
We extracted pedestrian trajectories from the whole setup of a ring experiment.
This data was transformed to a quasi-one-dimensional straight line.
The resulting fundamental diagram for the straight and curved part have the same shape and are comparable, tested with the Kolmogorov-Smirnov test.
The fundamental diagram of both parts shows a coexistence of two differing speed zones in specific density regions.
A stop speed was introduced to distinguish between moving and standing pedestrians.
Standing pedestrians define a stop wave that can be described by the characteristics length and duration.
While the maximum number of pedestrians in a stop wave in our data is increasing with growing density, the average length decreases.
In the future, we want to analyze those characteristics in a quantitative way.

\vspace*{-0.12cm}
\begin{acknowledgement}
Dedicated to the memory of Matthias Craesmeyer.

This study was performed within the project 'BaSiGo -- Bausteine f\"ur die Sicherheit von Gro\ss{}veranstaltungen' (Safety and Security Modules for Large Public Events), grant number: 13N12045, funded by the Federal Ministry of Education and Research (BMBF).
It is a part of the program on 'Research for Civil Security –- Protecting and Saving Human Life'.
\end{acknowledgement}

\vspace*{-0.2cm}
\bibliographystyle{spmpsci}
\bibliography{TGF_Ziemer_2015}

\end{document}